\documentstyle[aps,epsf,prl,epsfig,twocolumn]{revtex}
\def\be{\begin{equation}}
\def\ee{\end{equation}}
\def\bea{\begin{eqnarray}}          
\def\eea{\end{eqnarray}}
\def\bi{\begin{itemize}}
\def\ei{\end{itemize}}

\begin{document}

\title{ Dynamics of quantum phase transition: 
        exact solution in quantum Ising model  }

\author{ Jacek Dziarmaga }

\address{ Institute of Physics and Centre for Complex Systems,
          Jagiellonian University,
          Reymonta 4, 30-059 Krak\'ow, Poland  }

\date{ September 17, 2005 }

\maketitle

\begin{abstract}
Quantum Ising model is an exactly solvable model of quantum phase transition.
This paper gives an exact solution when the system is driven through the 
critical point at finite rate. The evolution goes through a series of Landau-Zener 
level anticrossings when pairs of quasiparticles with opposite pseudomomenta 
get excited with probability depending on the transition rate. Average density 
of defects excited in this way scales like a square root of the transition rate. 
This scaling is the same as the scaling obtained when the standard Kibble-Zurek 
mechanism of thermodynamic second order phase transitions is applied to the 
quantum phase transition in the Ising model.  
\end{abstract}


{\bf Introduction.---}
Phase transition is a fundamental change in a state of a system when
one of parameters of the system passes through its critical point. The 
states on the opposite sides of the critical point are characterized by 
different types of ordering. In a second order phase transition 
the fundamental change is continuous and the critical point is characterized 
by diverging correlation length and relaxation time. This critical slowing down 
implies that no matter how slowly a system is driven through the transition its 
evolution cannot be adiabatic close to the critical point. If it were adiabatic, 
then the system would continuously evolve between the two types of 
ordering. However, in the necessarily non-adiabatic evolution ordering of the 
state after the transition is not perfect, the state is a mosaic of ordered 
domains whose finite size depends on the rate of the transition. This scenario
was first described by Kibble \cite{KZ} and then its underlying dynamical mechanism 
was proposed by Zurek \cite{KZ} who predicted that the size of the ordered domains 
scales with the transition time $\tau_Q$ as $\tau_Q^w$, where $w$ is 
a combination of critical exponents. The Kibble-Zurek mechanism (KZM) of second
order thermodynamic phase transitions was confirmed by numerical
simulations of the time-dependent Ginzburg-Landau model \cite{KZnum} and
tested by experimens in liquid crystals \cite{LC}, superfluid helium 3 \cite{He3},
superfluid helium 4 \cite{He4}, and both high-$T_c$ \cite{highTc} and 
low-$T_c$ \cite{lowTc} superconductors. KZM is a universal theory of dynamics 
of second order phase transition whose applications range from the low temperature 
Bose-Einstein condensation (BEC) \cite{BEC} to the ultra high temperature transitions 
in the grand unified theories of high energy physics. However, the zero temperature 
quantum limit remains largely unexplored (but see 
Refs.\cite{3sites,Polkovnikov,KZIsing}) and 
quantum phase transitions are in many respects qualitatively different from transitions 
at finite temperature. Most importantly time evolution is unitary, there is no damping 
that seems to play essential role in KZM. 

According to Sachdev \cite{book} understanding of quantum phase transitions is based 
on two prototypical models. One is Bose-Hubbard model with its transition from Mott 
insulator to superfluid and the other is the 1-dimensional quantum Ising model. Of the 
two only the Ising model is exactly solvable. It is defined by the Hamiltonian 
\be
H~=~-J \sum_{n=1}^N \left( g\sigma^x_n + \sigma^z_n\sigma^z_{n+1} \right)~.
\label{Hsigma}
\ee 
with periodic boundary conditions 
\be
\vec\sigma_{N+1}~=~\vec\sigma_1~.
\label{periodicsigma}
\ee
The 1-dimensional quantum Ising model has the same critical properties as the 
2-dimensional classical Ising model. Quantum phase transition takes place at the 
critical value $g=1$ of external magnetic field. When $g\gg 1$, then the ground 
state is a paramagnet
$|\rightarrow\rightarrow\rightarrow\dots\rightarrow\rangle$ with all
spins polarized up along the $x$-axis. On the other hand when $g\ll 1$, then
there are two degenerate ferromagnetic ground states with all spins 
pointing either up or down along the $z$-axis: 
$|\uparrow\uparrow\uparrow\dots\uparrow\rangle$ or 
$|\downarrow\downarrow\downarrow\dots\downarrow\rangle$. In adiabatic
transition from paramagnet to ferromagnet the system would choose one of the
two ferromagnetic states (with possible help of infinitezimal symmetry breaking 
field along the $z$-axis, as usual in symmetry breaking phase transitions). However, 
when $N\to\infty$, then energy gap at $g=1$ tends to zero (quantum version of the 
critical slowing down) and it is 
impossible to pass the critical point without exciting the system.
As a result the system ends in a quantum superposition of states like
\be
|\dots
\uparrow
\downarrow\downarrow\downarrow\downarrow\downarrow
\uparrow\uparrow\uparrow\uparrow\uparrow\uparrow\uparrow
\downarrow\downarrow\downarrow\downarrow
\uparrow\uparrow\uparrow\uparrow\uparrow\uparrow
\downarrow
\dots\rangle
\label{domains}
\ee 
with finite domains of spins pointing up or down and separated by kinks where the 
polarization of spins changes its orientation. Average size of the domains or, 
equivalently, average density of kinks depends on transition rate. When 
transition is slow, then the size is large, but when it is very fast, then orientation 
of individual spins can become random, uncorrelated with their nearest neighbors. 
Transition time $\tau_Q$ can be unambiguously defined when we assume that close to 
the critical point at $g=1$ time-dependent field $g(t)$ driving the transition can 
be approximated by a linear quench 
\be
g(t<0)~=~-\frac{t}{\tau_Q}~
\label{linear}
\ee
with variable quench time $\tau_Q$. Density of kinks after the linear quench was 
estimated in Ref.\cite{KZIsing} as 
\be
n~\simeq~\left(\frac{\hbar}{2J\tau_Q}\right)^{1/2}~
\label{KZscaling}
\ee    
in an attempt to generalize KZM to quantum phase transitions. This is an order of 
magnitude estimate with an unknown ${\cal O}(1)$ prefactor. Results of numerical 
simulations in Ref.\cite{KZIsing} for a chain of $N$ spins were fitted with 
$n\sim\tau_Q^{-w}$. The fitted exponent $w$ decreased with increasing $N$ reaching 
$w=0.58$ for $N=100$. This is roughly consistent with the predicted $w=\frac12$. 

These encouriging results strongly motivate further study of the dynamics of quantum 
phase transition in the quantum Ising model. As this model is exactly solvable dynamics 
of its quantum phase transition also deserves to be solved exactly. In fact this is the 
only model, without any counterpart in the thermodynamic transitions, where 
KZM can be compared with an exact analytic solution. I begin by
reminding relevant facts about energy spectrum of the model. 

{\bf Energy spectrum.---}
Here I assume $N$ is even for convenience. After Jordan-Wigner transformation \cite{JW},
\bea
&&
\sigma^x_n~=~1-2 c^\dagger_n  c_n~, \\
&&
\sigma^z_n~=~
-\left( c_n+ c_n^\dagger\right)
 \prod_{m<n}(1-2 c^\dagger_m c_m)~,
\eea
introducing fermionic operators $c_n$ which satisfy anticommutation relations 
$\left\{c_m,c_n^\dagger\right\}=\delta_{mn}$ and 
$\left\{ c_m, c_n \right\}=\left\{c_m^\dagger,c_n^\dagger \right\}=0$
the Hamiltonian (\ref{Hsigma}) becomes \cite{LSM}
\be
 H~=~P^+~H^+~P^+~+~P^-~H^-~P^-~,
\label{Hc}
\ee
where
\be
P^{\pm}=
\frac12\left[1\pm\prod_{n=1}^N\sigma^x_n\right]=
\frac12\left[1~\pm~\prod_{n=1}^N\left(1-2c_n^\dagger c_n\right)\right]
\label{Ppm}
\ee
are projectors on the subspaces with even ($+$) and odd ($-$) numbers of 
$c$-quasiparticles and  
\bea
H^{\pm}~=~
J \sum_{n=1}^N
\left( 
g c_n^\dagger  c_n - c_n^\dagger  c_{n+1} - c_{n+1}  c_n - \frac{g}{2} +{\rm h.c.} 
\right)~.
\label{Hpm}
\eea
are corresponding reduced Hamiltonians. The $c_n$'s in $H^-$ satisfy
periodic boundary conditions $c_{N+1}=c_1$, but the $c_n$'s in $H^+$
must obey $c_{N+1}=-c_1$ what I call ``antiperiodic'' boudary conditions. 

The parity of the number of $c$-quasiparticles is a good quantum number and the ground
state has even parity for any value of $g$. Assuming that a quench begins in the ground 
state we can confine to the subspace of even parity. $H^+$ is diagonalized by Fourier 
transform followed by Bogoliubov transformation \cite{LSM}. Fourier transform consistent 
with the antiperiodic boundary condition $c_{N+1}=-c_1$  
\be
c_n~=~ 
\frac{e^{-i\pi/4}}{\sqrt{N}}
\sum_k c_k e^{ik(na)}~,
\label{Fourier}
\ee
where the pseudomomenta $k$ take ``half integer'' values
\be
k~=~
\pm \frac12 \frac{2\pi}{Na},
\dots,
\pm \left(\frac{N}{2}-\frac12\right) \frac{2\pi}{Na}~,
\label{k}
\ee
transforms the Hamiltonian into
\bea
&&
H^+~=\nonumber\\
&&
J\sum_k
\left\{
2[g-\cos(ka)]  c_k^\dagger  c_k +
\sin(ka)
\left[ 
 c^\dagger_k  c^\dagger_{-k}+
 c_{-k}  c_k
\right]
-g
\right\}~.
\label{Hck}
\eea
Here $a$ is lattice spacing. Diagonalization of $H^+$ is completed by 
the Bogoliubov transformation
\be
c_k~=~
u_k  \gamma_k + v_{-k}^*  \gamma^\dagger_{-k}~
\label{Bog}
\ee
provided that Bogoliubov modes $(u_k,v_k)$ are eigenstates of the stationary
Bogoliubov-de Gennes equations
\bea
\epsilon~ u_k &=& +2J[g-\cos(ka)] u_k+2J\sin(ka) v_k~,\nonumber\\
\epsilon~ v_k &=& -2J[g-\cos(ka)] v_k+2J\sin(ka) u_k~ \label{stBdG}
\eea
There are two eigestates for each $k$ with eigenenergies $\epsilon=\pm\epsilon_k$,
where  
\be
\epsilon_k~=~2J\sqrt{[g-\cos(ka)]^2+\sin^2(ka)}~.
\ee
The positive energy eigenstate 
\be
(u_k,v_k)\sim\left[(g-\cos ka)+\sqrt{g^2-2g\cos ka+1},\sin(ka)\right]~,
\label{uvplus}
\ee
normalized so that $|u_k|^2+|v_k|^2=1$, defines the quasiparticle operator 
$\gamma_k~=~u_k^* c_k + v_{-k} c_{-k}^\dagger$,
and the negative energy eigenstate 
$
(u^-_k,v^-_k)=(-v_k,u_k)~ 
\label{uv-}
$
defines $\gamma_k^-=(u_k^-)^*c_k+v_{-k}^-c_{-k}^\dagger=-\gamma_{-k}^\dagger$. 
After the Bogoliubov transformation the Hamiltonian 
$
H^+=
\frac12
\sum_k \epsilon_k
\left( 
\gamma_k^\dagger \gamma_k - \gamma_k^{-\dagger} \gamma_k^-    
\right)
$
equivalent to
\be
H^+~=~
\sum_k \epsilon_k~
\left(
\gamma_k^\dagger \gamma_k-\frac12
\right)~
\label{Hgamma}
\ee
which is a simple-looking sum of quasiparticles with half-integer pseudomomenta. 
However, thanks to the projection $P^+~H^+~P^+$ in Eq.(\ref{Hc}) only states 
with even numbers of quasiparticles belong to the spectrum of $H$.   

{\bf Transition from paramagnet to ferromagnet. ---}
In the linear quench Eq.(\ref{linear}) the system is initially prepared in its 
ground state at large initial value of $g\gg 1$ but when $g$ is ramped down to 
zero the state of the system gets excited from its instantaneous ground state
and in general its final state at $t=0$ has finite number of kinks. Comparing 
the Ising 
Hamiltonian Eq.(\ref{Hsigma}) at $g=0$ with the Bogoliubov Hamiltonian (\ref{Hgamma}) 
at $g=0$ we obtain a simple expression for the operator of the number of kinks 
\bea
{\cal N} ~\equiv~ \frac12 \sum_{n=1}^{N} 
                  \left(1-\sigma^z_n\sigma^z_{n+1}\right)~=~
                  \sum_k \gamma_k^\dagger \gamma_k~.   
\label{calN}
\eea 
The number of kinks is equal to the number of quasiparticles excited at $g=0$. 
The excitation probability 
\be
p_k~=~\langle\psi(0)|\gamma_k^\dagger\gamma_k|\psi(0)\rangle~
\ee
in the final state can be found with the time-dependent Bogoliubov method. 

The initial ground state is Bogoliubov vacuum $|0\rangle$ annihilated by all 
quasiparticle operators $\gamma_k$ which are determined by the asymptotic form 
of the (positive energy) Bogoliubov modes $(u_k,v_k)\approx(1,0)$ in the regime 
of $g\gg1$. When $g(t)$ is ramped down, then the quantum state $|\psi(t)\rangle$
in general gets excited from the instantaneous ground state. The time-dependent 
Bogoliubov method makes an Ansatz that $|\psi(t)\rangle$ is Bogoliubov vacuum 
annihilated by a set of quasiparticle annihilation operators $\tilde{\gamma}_k$ 
defined by a time-dependent Bogoliubov transformation
\be
c_k ~=~ u_k(t)  \tilde{\gamma}_k + v_{-k}^*(t)  \tilde{\gamma}^\dagger_{-k}~
\label{tildeBog}
\ee
with the initial condition $[u_k(-\infty),v_k(-\infty)]=(1,0)$. The Bogoliubov 
modes $[u_k(t),v_k(t)]$ must satisfy Heisenberg equation
$
i\hbar\frac{d}{dt} c_k~=~\left[ c_k, H^+\right]
$
with the constraint $\frac{d}{dt}\tilde{\gamma}_k=0$ [equivalent to 
$\tilde{\gamma}_k|\psi(t)\rangle=0$]. The Heisenberg equation is equivalent to 
the dynamical version of the Bogoliubov-de Gennes equations (\ref{stBdG}):
\bea
i\hbar\frac{d}{dt} u_k &=&
+2J\left[g(t)-\cos(ka)\right] u_k +
 2J \sin(ka)~ v_k~,\nonumber\\
i\hbar\frac{d}{dt} v_k &=&
-2J\left[g(t)-\cos(ka)\right] v_k +
 2J \sin(ka)~ u_k~.
\label{dynBdG} 
\eea
At any value of $g$ Eqs.(\ref{dynBdG}) have two instantaneous eigenstates. 
Initially the mode $[u_k(t),v_k(t)]$ is the positive energy eigenstate, but during 
the quench it gets ``excited'' to a combination of the positive and negative mode. 
At the end of the quench at $t=0$ when $g=0$ we have
\be
\left[u_k(0),v_k(0)\right]=
\alpha_k
\left(u_k,v_k\right)+
\beta_k
\left(u_k^-,v_k^-\right)
\ee
and consequently $\tilde{\gamma}_k=\alpha_k\gamma_k-\beta_k\gamma_k^\dagger$.
The final state which is, by construction, annihilated by both $\tilde{\gamma}_k$ and 
$\tilde{\gamma}_{-k}$ is
\be
|\psi(0)\rangle=
\prod_{k>0}
\left(
\alpha_k+\beta_k\gamma^\dagger_k\gamma^\dagger_{-k}
\right)
|0\rangle~.
\label{finalstate}
\ee
Pairs of quasiparticles with pseudomomenta $(k,-k)$ are excited with probability
\be
p_k~=~|\beta_k|^2~
\label{pbeta}
\ee
which can be found by mapping Eqs.(\ref{dynBdG}) to the Landau-Zener (LZ) problem 
(similarity between KZM and LZ problem was first pointed out by Damski in 
Ref.\cite{Bodzio}). The transformation 
\be
\tau~=~4J\tau_Q\sin(ka)\left(\frac{t}{\tau_Q}+\cos(ka)\right)
\label{tau}
\ee
brings Eqs.(\ref{dynBdG}) to the standard LZ form \cite{LZF}
\bea
i\hbar\frac{d}{d\tau} u_k &=&
-\frac12(\tau\Delta_k) u_k + \frac12 v_k ~,\nonumber\\
i\hbar\frac{d}{d\tau} v_k &=&
+\frac12(\tau\Delta_k) v_k + \frac12 u_k ~,\label{LZ}
\label{BdGLZ}
\eea
with $\Delta_k^{-1}=4J\tau_Q\sin^2(ka)$. Here the time $\tau$ runs from $-\infty$ to  
$\tau_{\rm final}=2J\tau_Q\sin(2ka)$ corresponding to $t=0$. Tunnelling between the 
positive and negative energy eigenstates happens when 
$\tau\in(-\Delta_k^{-1},\Delta_k^{-1})$. $\tau_{\rm final}$ is well outside this 
interval, $\tau_{\rm final}\gg\Delta_k^{-1}$, for 
long wavelength modes with $|ka|\ll\frac{\pi}{4}$. For these modes time $\tau$ 
in Eqs.(\ref{LZ}) can be extended to $+\infty$ making them fully equivalent to 
LZ equations \cite{LZF}. 

In slow transitions only long wavelength modes can get excited. For these modes 
we can use the LZ formula \cite{LZF} for excitation probability:
\be
p_k~\simeq~
e^{-\frac{\pi}{2\hbar\Delta_k}}~\approx~
e^{-2\pi \left(J\tau_Q/\hbar\right) (ka)^2}~.
\label{LZpk}
\ee 
This calculation is self-consistent when the width of the obtained
gaussian $ka=(4\pi J\tau_Q/\hbar)^{-1/2}$ is much less than $\frac{\pi}{4}$
or, equivalently, for slow enough quenches with $\tau_Q\gg\frac{4\hbar}{\pi^3J}$.
With the LZ formula (\ref{LZpk}) we can calculate the number of kinks 
in Eq.(\ref{calN}) as
\be
{\cal N}~=~\sum_k~p_k~.
\label{LZn}
\ee
There are at least two interesting special cases.

1) When $N\to\infty$ i.e. in the limit of proper phase transition the sum in 
Eq.(\ref{LZn}) can be replaced by an integral. Expectation value of density of 
kinks becomes
\be
n=\lim_{N\to\infty}\frac{\cal N}{N}=
\frac{1}{2\pi}\int_{-\pi}^{\pi}d(ka)~p_k=
\frac{1}{2\pi}
\frac{1}{\sqrt{2J\tau_Q/\hbar}}.
\label{scaling}
\ee    
The density scales like $\tau_Q^{-1/2}$ in agreement with KZM. It is a 
factor of $\frac{1}{2\pi}\approx 0.159$ less than the simple estimate in 
Eq.(\ref{KZscaling}) based on KZM \cite{KZIsing}. The results of numerical 
simulations in Ref.\cite{KZIsing} are best fitted with a prefactor of $0.16$ 
which is consistent with the exact number. 

2) For a finite chain we can ask what is the fastest $\tau_Q$ when still no 
kinks get excited. This critical $\tau_Q^{\rm ad}$ marks a crossover between 
adiabatic and non-adiabatic regimes. In other words we can ask what is the 
probability for a finite chain to stay in the ground state. As different pairs 
of quasiparticles $(k,-k)$ evolve independently, the probability to stay in the 
ground state is the product
\be
{\cal P}_{\rm GS}~=~
\prod_{k>0}\left(1-p_k\right)~.
\label{calP}
\ee
Well on the adiabatic side only the pair $\left(\frac{\pi}{N},-\frac{\pi}{N}\right)$ 
is likely to get excited and we can approximate
\be
{\cal P}_{\rm GS}~\approx~
1-p_{\frac{\pi}{N}}~\approx~
1-\exp\left(-2\pi^3\frac{J\tau_Q}{\hbar N^2}\right)~.
\label{calPapprox}
\ee
Again, numerical results in Ref.\cite{KZIsing} are consistent with
this formula: the best fit gives $59$ in place of the $2\pi^3\approx 62$.
A quench in a finite chain becomes non-adiabatic when $\tau_Q$ is less than
$
\tau_Q^{\rm ad}=\frac{\hbar N^2}{2\pi^3 J} 
$
which grows with the system size like $N^2$. In other words,
the size $N$ of a defect-free chain grows like $\tau_Q^{1/2}$ 
in consistency with Eq.(\ref{scaling}).
 
{\bf Transition from ferromagnet to paramagnet.---}
Again, close to the critical point at $g=1$ the external field can be approximated
by a linear quench $g(t>0)=\frac{t}{\tau_Q}$. The initial state is the
even parity ground state at $g=0$ and the final state $|\psi(t)\rangle$
is in general excited with respect to the polarized ground state at a large 
final $g\gg 1$. We want to know how many spins in the final state are flipped
with respect to the polarization in the ground state or, more precisely, what is 
the expectation value of the operator of the number of flips 
${\cal F}=\frac12\left(N-\sum_{n=1}^N\sigma^x_n\right)$. We note that when $g\gg 1$, 
then $H\approx-Jg\sum_{n=1}^N\sigma^x_n=-Jg(N-2{\cal F})$, but on the other hand
$H^+\approx 2Jg\sum_k\hat\gamma_k^\dagger\hat\gamma_k-JgN$, and consequently
\be
{\cal F}~=~\sum_k\hat\gamma_k^\dagger\hat\gamma_k~.
\label{calF}
\ee
At large $g\gg 1$ the number of flips is the number of excited quasiparticles.
Formal similarity of Eqs.(\ref{calF}) and (\ref{calN}) allows us to proceed in the
same way as in the paramagnet to ferromagnet transition. We soon arrive at 
Eqs.(\ref{BdGLZ}) but with $\tau$ replaced by $-\tau$ on the right hand side.
In analogy to Eq.(\ref{scaling}) we find density of flipped spins
\be
f~=~
\frac{1}{2\pi}~
\frac{1}{\sqrt{2J\tau_Q/\hbar}}~
\label{Fscaling}
\ee    
and the probability for a finite chain to stay in the ground state is again
given by Eq.(\ref{calP}).

{\bf Conclusion.---}
This paper gives an exact analytic solution when the quantum Ising model is 
driven through 
its quantum critical point at finite transition rate. The evolution goes through 
a series of Landau-Zener level anticrossings when pairs of quasiparticles with 
opposite pseudomomenta get excited with probability depending on the transition 
rate. Average density of defects excited in this way scales like a square root of
the transition rate. This scaling is the same as the scaling obtained when the 
standard Kibble-Zurek mechanism of thermodynamic second order phase transitions 
is extended to the quantum phase transition in the Ising model.  

{\bf Acknowledgements. --- } 
I would like to thank Bogdan Damski and Wojtek Zurek for stimulating
discussions and Marek Rams for critical reading of the manuscript. 
This work was supported in part by Polish Government scientific funds (2005-2008) 
as a research project and in part by ESF COSLAB program.
 

\end{document}